\documentclass[twocolumn,showpacs,showkeys,preprintnumbers,amsmath,amssymb,prl]{revtex4}

\usepackage{graphicx}
\usepackage{dcolumn}
\usepackage{bm}

\begin{document}

\title{$ {\bf VO_2} $: A Novel View from Band Theory}

\author{V.\ Eyert}
\altaffiliation[Present address: ]{Materials Design, 92120 Montrouge, France} 
\email[Email: ]{veyert@materialsdesign.com} 
\affiliation{Center for Electronic Correlations and Magnetism, 
             Institute of Physics, University of Augsburg,
             86135 Augsburg, Germany}

\date{\today}

\begin{abstract}
New calculations for vanadium dioxide, one of the most controversely 
discussed materials for decades, reveal that band theory as based on 
density functional theory is well capable of correctly describing the 
electronic and magnetic properties of the metallic as well as both 
the insulating $ {\rm M_1} $ and $ {\rm M_2} $ phases. Considerable 
progress in the understanding of the physics of $ {\rm VO_2} $ is 
achieved by the use of the recently developed hybrid functionals, 
which include part of the electron-electron interaction exactly and 
thereby improve on the weaknesses of semilocal exchange functionals as 
provided by the local density and generalized gradient approximations. 
Much better agreement with photoemission data as compared to previous 
calculations is found and a consistent description of the rutile-type 
early transition-metal dioxides is achieved. 
\end{abstract}


\maketitle


The metal-insulator transition (MIT) of stoichiometric $ {\rm VO_2} $ 
at ambient pressure \cite{morin59} has been a matter of ongoing 
controversy for decades. The issue attained its fundamental character 
from the discovery of a structural distortion occurring at the same 
temperature (340\,K) as the MIT \cite{andersson54,goodenough60}. 
Since then, discussion has focused on resolving the hen-and-egg problem 
of identifying the driving force of the combined electronic-structural 
transition. The latter is from the high-temperature rutile phase to the 
low-temperature monoclinic, so-called $ {\rm M_1} $ phase. The distortions 
occurring at the transition are characterized by metal dimerization 
parallel to the rutile $ c $ axis and a zigzag-like antiferroelectric 
shift of the V atoms perpendicular to this axis out of the center 
of the surrounding O octahedra. In a simple molecular-orbital picture 
these structural changes affect mainly the V $ 3d $ $ t_{2g} $-derived 
states, which in the rutile phase straddle the Fermi energy and have 
similar band occupations \cite{goodenough60,goodenough71}. Yet, due to 
the peculiarities of this structure with its characteristic octahedral 
chains parallel to the $ c $ axis, the $ t_{2g} $ states naturally fall 
into the $ d_{\parallel} $ orbitals and the so-called $ \pi^{\ast} $ 
orbitals, which show a quasi-onedimensional and rather threedimensional 
dispersion, respectively \cite{vo2rev}. Response to the structural 
changes occurring at the transition to the $ {\rm M_1} $ phase is 
diverse. While the $ d_{\parallel} $ bands split into occupied bonding 
and empty antibonding states, the $ \pi^{\ast} $ states experience 
energetical upshift leading to their depopulation 
\cite{goodenough60,goodenough71}. As a consequence, a bandgap opens 
between the bonding $ d_{\parallel} $ and the $ \pi^{\ast} $ bands. 
An alternative interpretation, emphasizing the electronic origin of 
the MIT, assigns the splitting of the $ d_{\parallel} $ bands in the 
insulating phase rather to strong electronic correlations, which are 
screened by the $ \pi^{\ast} $ electrons in the metallic phase 
\cite{zylbersztejn75}.

Still, the matter was complicated by the discovery of two additional 
insulating phases, namely the monoclinic $ {\rm M_2} $ and the 
triclinic $ {\rm T} $ phase occurring in doped samples or under 
uniaxial pressure, which suppresses the zigzag-like shift in half of 
the octahedral chains \cite{marezio72}. In the $ {\rm M_2} $ phase, 
these chains still show metal dimerization, while in the remaining 
chains the dimerization is lost in favor of quasi-onedimensional 
antiferromagnetic ordering. The T phase is intermediate between the 
two monoclinic phases as, compared to the $ {\rm M_2} $ phase, the 
dimerized chains start to zigzag and the zigzag chains start to 
dimerize until both chains are identical and the $ {\rm M_1} $ phase 
is entered. Of course, the observation of the $ {\rm M_2} $ phase 
reinitiated the debate about the driving force of the phase transitions 
and questioned especially the interpretation in terms of particular 
structural changes underlying the molecular orbital picture.

A somewhat related but more academic aspect of the debate is the question 
as to what extent band theory would be able to correctly describe and 
understand the MIT and the role of electronic correlations. This discussion 
has its roots in the spectacular failure of calculations based on 
density functional theory (DFT) within the local density (LDA) or 
generalized gradient approximation (GGA) to reproduce a finite band gap in 
the single-particle spectrum obtained from the monoclinic structures 
\cite{wentz94a,vo2rev}. Furthermore, from the existence of the 
$ S = \frac{1}{2} $-Heisenberg chains characterizing the $ {\rm M_2} $ 
phase as well as the fact that, due to simple electrostatic arguments 
connecting the pairing on one chain to the antiferroelectric shift on 
the neighboring chains, the $ {\rm M_1} $ and T phases may be regarded 
as superpositions of two $ {\rm M_2} $-type distortion patterns with equal 
and unequal amplitude, respectively, Rice {\em et al.}\ concluded, that 
all insulating phases of $ {\rm VO_2} $ must be of the Mott-Hubbard 
type and are not accessible by band theory \cite{rice94}. Indeed, LDA 
calculations for the $ {\rm M_2} $ phase, while leading to an 
antiferromagnetic ground state, missed the insulating gap like for the 
$ {\rm M_1} $ phase \cite{vo2rev}. This failure could be cured by 
LDA+$ U $ calculations, which, however, resulted in antiferromagnetic 
insulating ground states also for the $ {\rm M_1} $ and the metallic 
rutile phase \cite{korotin02}. So far, these latter two phases could 
be correctly reproduced only by GW calculations and combinations of 
LDA with the dynamic mean-field theory 
\cite{continenza99,gatti07,biermann05,tomczak07}. 
Nevertheless, a correct consistent description of the metallic rutile 
as well as both insulating monoclinic phases using a parameter-free 
methodology has not yet emerged.

Here, the results of renewed calculations as based on DFT are presented. 
In contrast to previous work the present investigation builds on the use 
of the recently developed class of hybrid functionals, which have already 
proven to allow for a strikingly improved description of semiconductors 
and insulators \cite{marsman08,shimazaki10}. Yet, applications of these 
new tools to materials, which are regarded as strongly correlated, are 
still rare due to the high computational demand \cite{marsman08}. The 
present work demonstrates that density functional theory in combination 
with hybrid functionals is well capable to generate  
an insulating gap in the single-particle spectrum of both the 
$ {\rm M_1} $ and $ {\rm M_2} $ phases and to properly describe the 
antiferromagnetic ordering on the Heisenberg chains of the latter. 
These results shed new light onto the discussion of the role of the 
relevant correlations, since hybrid functionals primarily aim at 
improving on the exchange part of the LDA/GGA, while leaving 
the correlation functional unaltered.


The calculations were performed using density functional theory as 
implemented in the ab-initio total-energy and molecular-dynamics 
program VASP (Vienna ab-initio simulation program) developed at 
the Universit\"at Wien \cite{vasp}. The exchange-correlation 
functional was considered at the level of the generalized gradient 
approximation (GGA) \cite{perdew96a}. In addition, calculations as 
based on the recently developed hybrid functonals were performed. 
Within the framework of the generalized Kohn-Sham scheme \cite{seidl96} 
these functionals combine the exchange functional as arising from 
the local density approximation (LDA) with the non-local Hartree-Fock 
expression. In the present work, the functional proposed by Heyd, 
Scuseria, and Ernzerhof (HSE) was used \cite{hse}. 
In this approach, the short-range 
part of the exchange functional is represented by a (fixed) combination 
of GGA- and Hartree-Fock contributions, while the long-range part and 
the correlation functional are described by the GGA only. 
The single-particle equations were solved using the projector-augmented 
wave (PAW) method \cite{paw,vasppaw} with a plane-wave basis with a 
cutoff of 400\,eV. 

\begin{figure}[tb]
\centering
\includegraphics[width=0.48\textwidth]{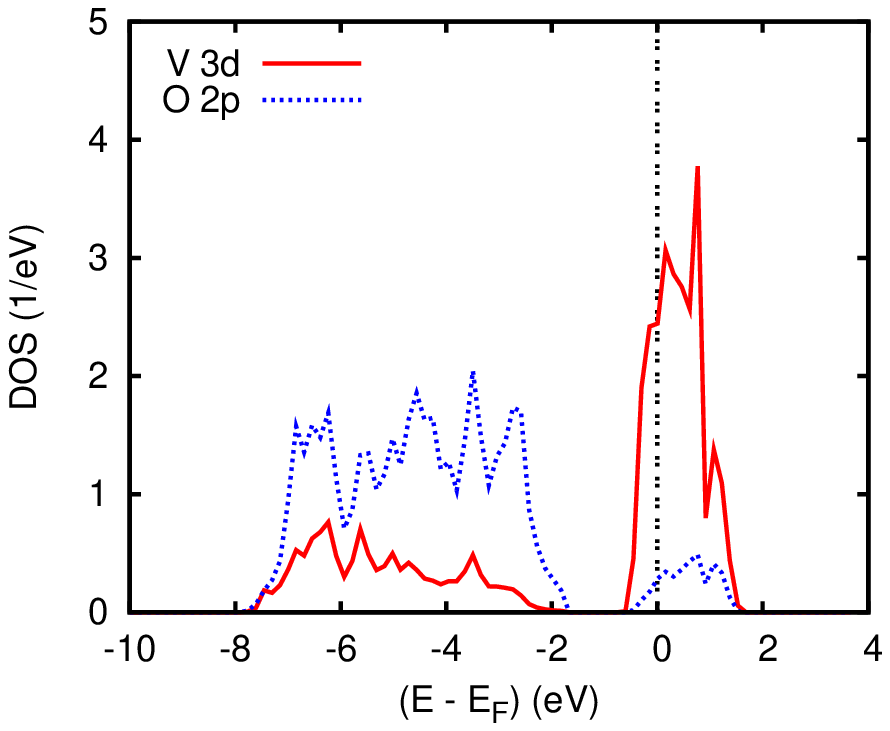}
\includegraphics[width=0.48\textwidth]{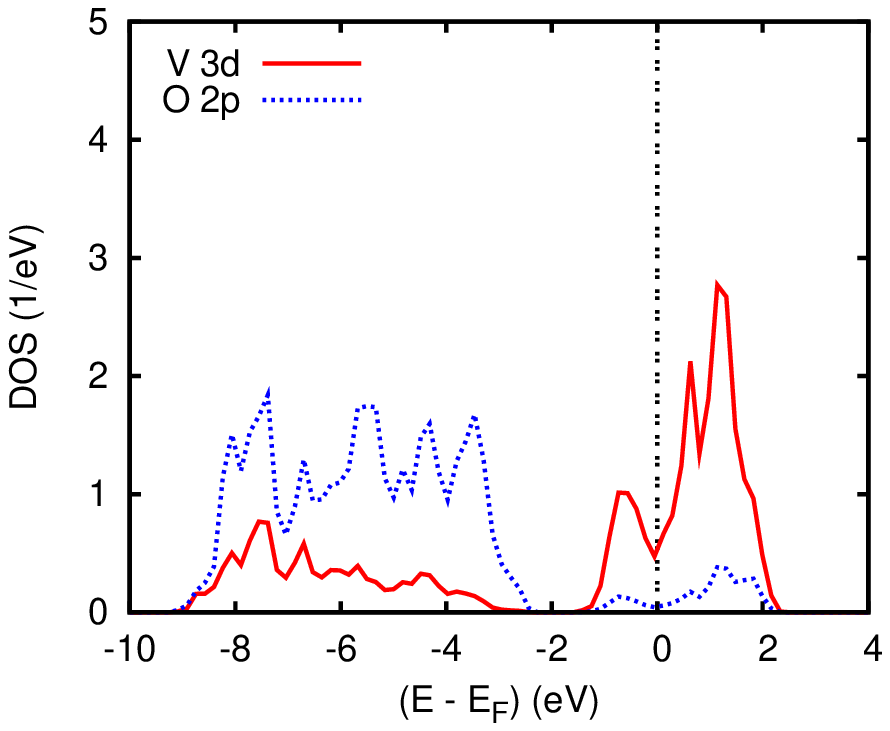}
\caption{(Color online) 
         Partial DOS of rutile $ {\rm VO_2} $ as calculated using 
         the GGA (top) and the HSE (bottom) functional. 
         }
\label{fig:res1}
\end{figure}

In a first step, the metallic rutile phase was considered. Structural 
data were taken from the work of McWhan {\em et al.}\ \cite{mcwhan74} in 
order to be consistent with previous calculations \cite{vo2rev}. 
Partial densities of states as emerging from spin-degenerate 
calculations are displayed in Fig.~\ref{fig:res1}. 
Two groups of bands are recognized. While the fully occupied bands 
derive mainly from the O $ 2p $ states, the group of 
bands straddling the Fermi energy are dominated by the V $ 3d $ states. 
V $ 3d $- and O $ 2p $-contributions showing up in the energy range, 
where the respective other orbitals dominate, result from hybridization 
between these states. The nearly perfect octahedral crystal field 
splits the V $ 3d $ levels into weakly $ \pi $-bonding 
$ t_{2g} $ states ranging from $ -0.8 $ to $ +1.8 $\,eV and strong 
$ \sigma $-bonding $ e_g $ bands, which are found at higher energies 
but not shown here. The GGA results as displayed in the upper panel 
of Fig.~\ref{fig:res1} are very similar to the partial densities of 
states as arising from previous LDA calculations \cite{wentz94a,vo2rev}. 

The HSE results given in the lower panel of Fig.~\ref{fig:res1} 
deviate substantially from those obtained by the GGA. While retaining 
their  shape, the O $ 2p $ partial densities of states experience 10\% 
broadening and energetical downshift by about 1\,eV. In contrast, the 
width of the V $ 3d $ states almost doubles leading to an occupied 
band width of about 1.5\,eV and a pronounced peak near $ -0.75 $\,eV in 
much better agreement with experiment than the GGA results. In particular,  
photoemission finds an occupied band width of about 1.5\,eV for the 
V $ 3d $ band centered at about $ -1 $\,eV \cite{shin90,okazaki04}. 

\begin{figure}[tb]
\centering
\includegraphics[width=0.48\textwidth]{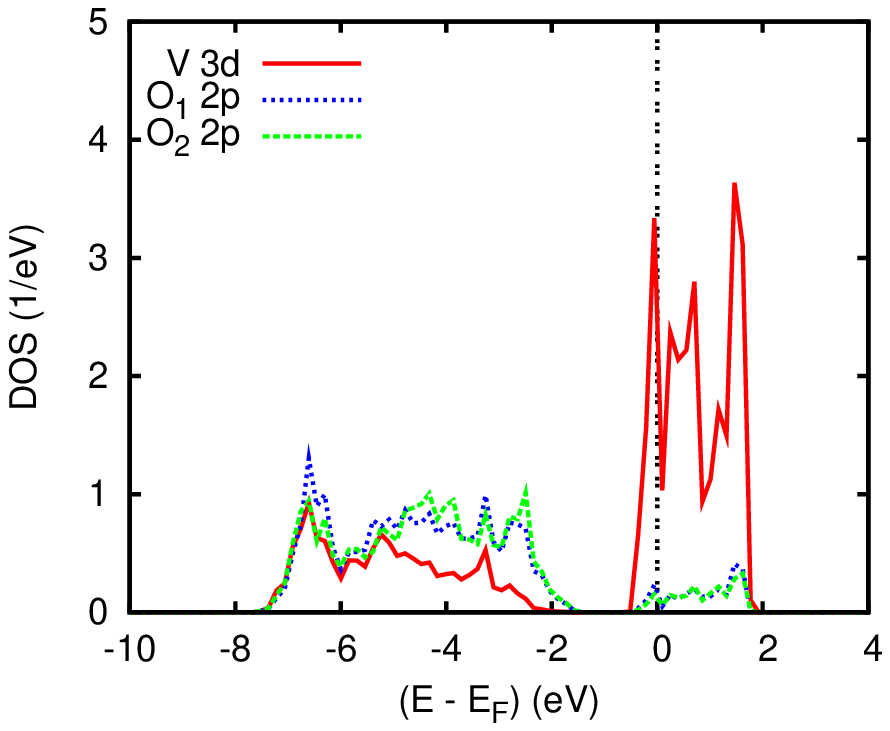}
\includegraphics[width=0.48\textwidth]{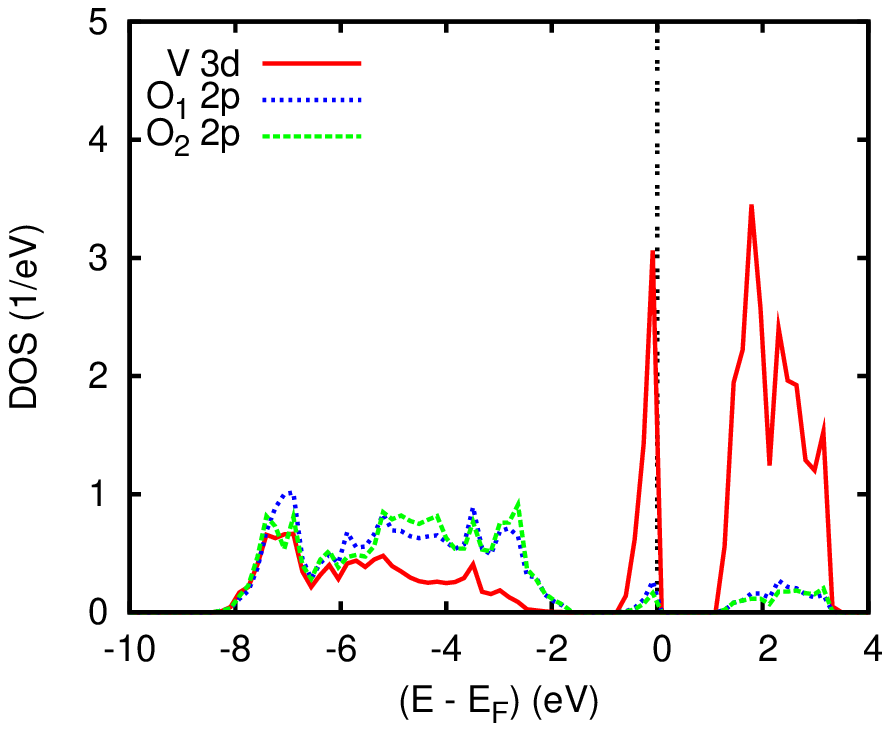}
\caption{(Color online) 
         Partial DOS of $ {\rm M_1} $-$ {\rm VO_2} $ as calculated using 
         the GGA (top) and the HSE (bottom) functional.} 
\label{fig:res2}
\end{figure}

\begin{figure}[tb]
\centering
\includegraphics[width=0.40\textwidth]{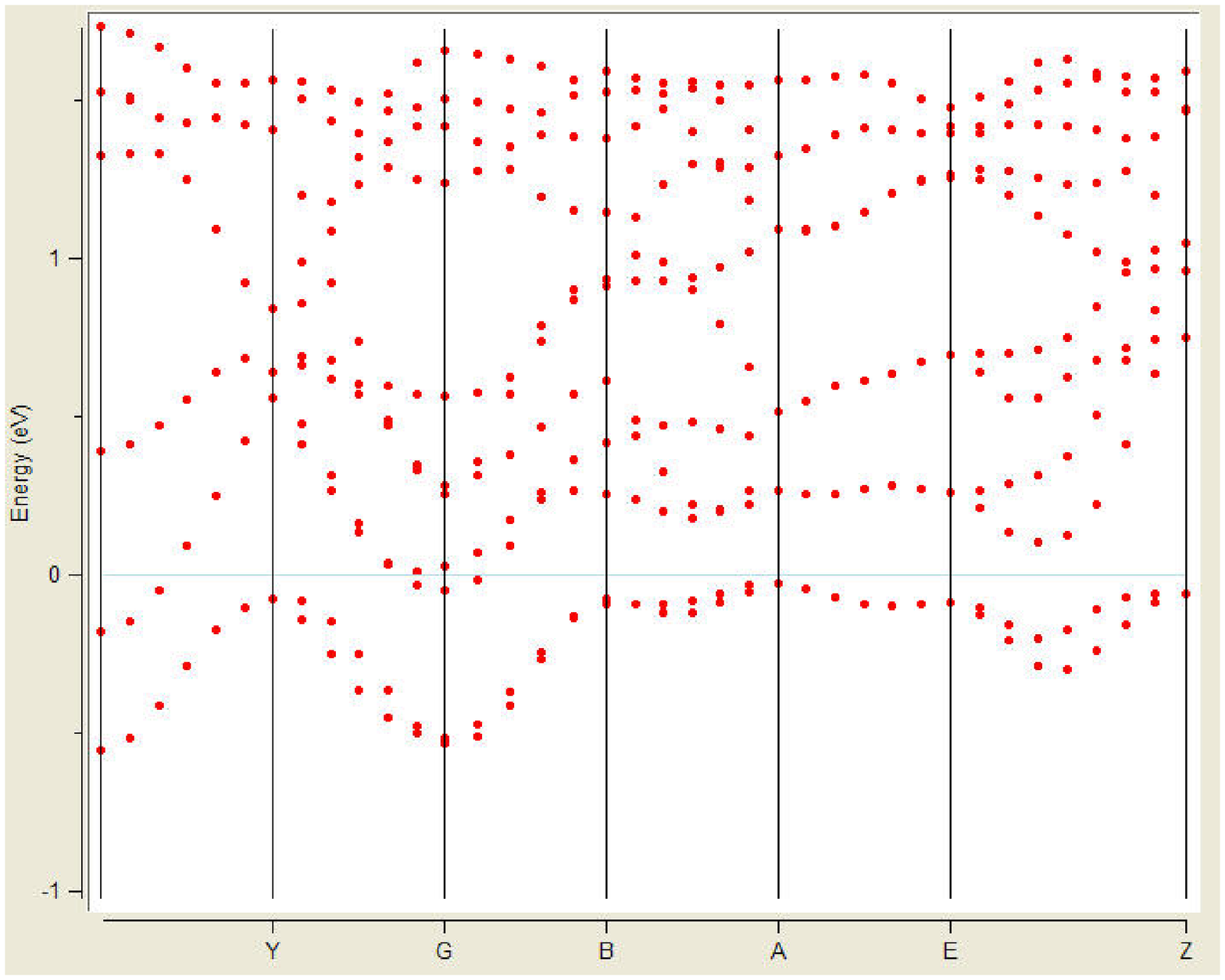}
\includegraphics[width=0.40\textwidth]{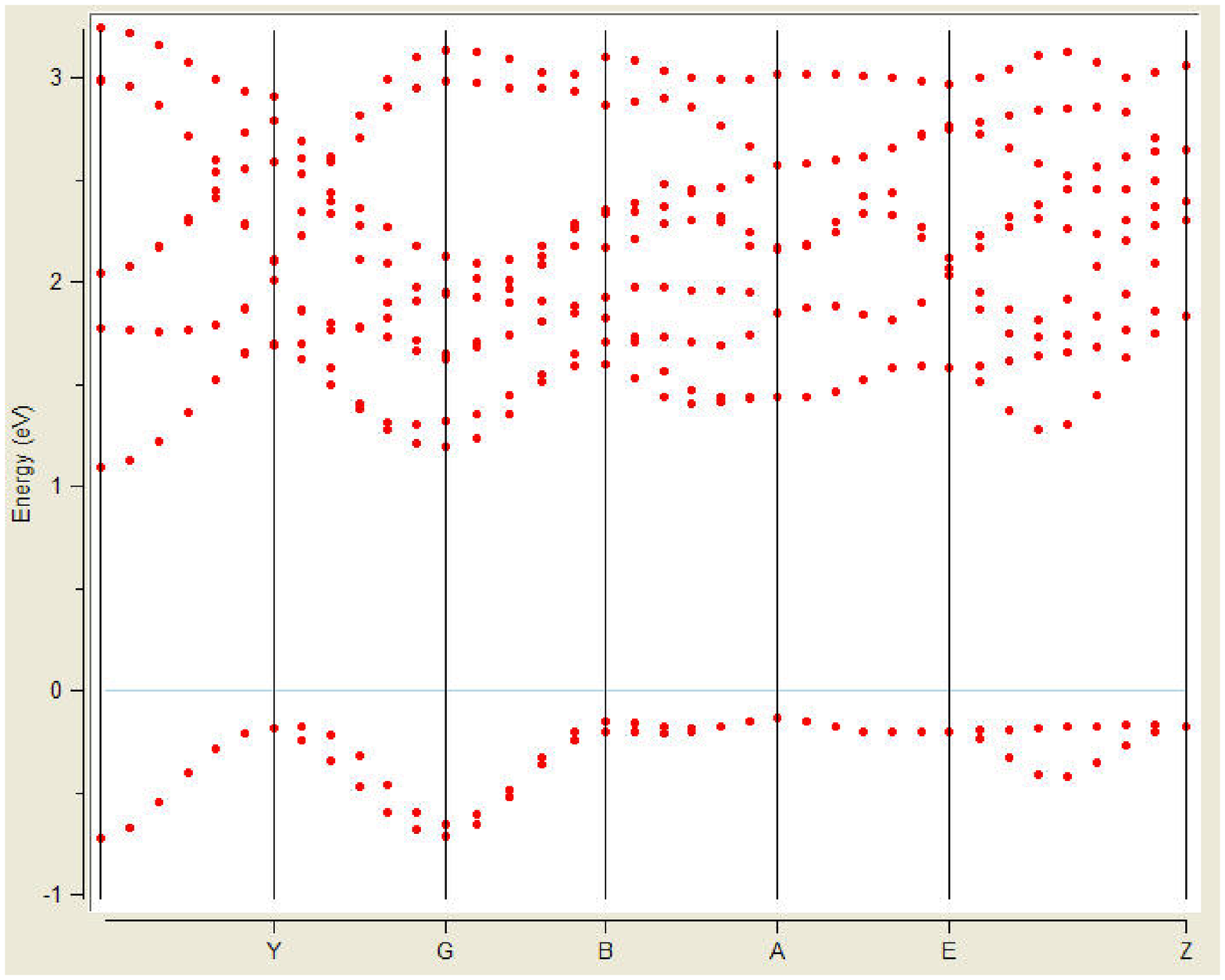}
\caption{(Color online) 
         Electronic bands of $ {\rm M_1} $-$ {\rm VO_2} $ as calculated 
         using the GGA (top) and the HSE (bottom) functional.} 
\label{fig:res2a}
\end{figure}

For the monoclinic $ {\rm M_1} $ phase, crystal structure data 
by Longo and Kierkegaard were used \cite{longo70}. Partial densities of 
states as arising from GGA calculations are shown in the upper panel 
of Fig.~\ref{fig:res2}.
As for the rutile structure, these results closely resemble those 
of previous calculations \cite{wentz94a,vo2rev}. Again, the bands 
fall into O $ 2p $ and V $ 3d $ dominated groups well below and 
at the Fermi energy. Although the V $ 3d $ states display a 
sharp dip at $ {\rm E_F} $, the calculations still reflect the 
above mentioned failure of LDA/GGA to reproduce the insulating 
band gap of the monoclinic phase. According to the band structure 
displayed in the upper panel of Fig.~\ref{fig:res2a} 
the metallic behaviour results from small semimetallic-like overlap of 
the characteristic double band ranging from $ -0.5 $\,eV to slightly 
above the Fermi energy with the almost empty higher lying bands. Yet, 
as has been discussed in detail previously, strong orbital ordering is 
obtained, since the occupied double band derives mainly from the 
$ d_{\parallel} $ states, whereas the $ \pi^{\ast} $ states are found 
mostly above $ {\rm E_F} $ \cite{vo2rev}. This is in striking contrast 
to the rutile phase, where LDA calculations yield very similar 
occupations of the $ t_{2g} $ bands with only small hybridization 
between the $ d_{\parallel} $ and $ \pi^{\ast} $ states \cite{vo2rev}. 
Thus, the MIT is accompanied by strong orbital switching, as has been 
confirmed by most calculations as well as by recent XAS measurements 
\cite{haverkort05}. 

The situation changes drastically on turning to the hybrid functional 
calculations. According to the partial densities of states displayed 
in the lower panel of Fig.~\ref{fig:res2}, the previously slightly 
overlapping groups of bands are pulled apart and an insulating band gap 
of $ \approx 1.1 $\,eV is opened between the bonding $ d_{\parallel} $ 
bands and the empty $ \pi^{\ast} $ states. The bonding $ d_{\parallel} $ 
bands form a sharp peak at about $ 0.75 $\,eV below the center of 
the gap, again in good agreement with photoemission experiments, which 
find the V $ 3d $ band of about 1\,eV width at a binding energy of 
about 1\,eV \cite{shin90,okazaki04,koethe06}. The band structure shown 
in the lower panel of Fig.\ \ref{fig:res2a} reveals a clear separation 
into fully occupied and empty bands. Furthermore, we note the striking 
similarity to the GGA bands as concerns both widths and shapes, the main 
difference being the rather rigid band shift.

\begin{figure}[tb]
\centering
\includegraphics[width=0.40\textwidth]{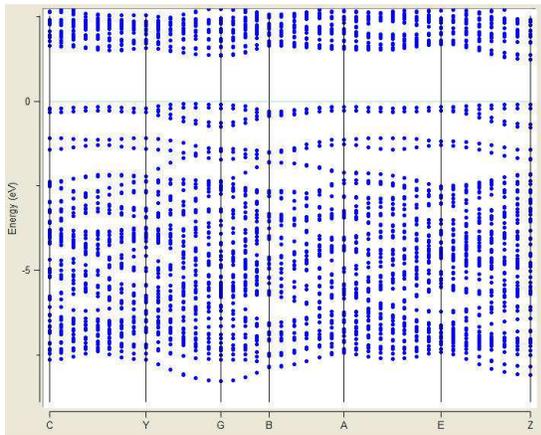}
\caption{(Color online) 
         Electronic bands of $ {\rm M_2} $-$ {\rm VO_2} $ as calculated 
         using the HSE functional.} 
\label{fig:res3}
\end{figure}

Finally, spin-polarized antiferromagnetic calculations were performed 
for the $ {\rm M_2} $ phase. Here, crystal structure data as given by 
Marezio {\em et al.}\ were used \cite{marezio72}. While previous LDA 
calculations resulted in a stable antiferromagnetic ground state 
with local magnetic moments of $ \approx 0.5 \mu_{\rm B} $ mainly 
carried by the V $ 3d $ $ d_{\parallel} $ states, the insulating 
behaviour could not be reproduced \cite{vo2rev}. In contrast, the 
present HSE calculations lead to V magnetic moments of 
$ 1.0 \mu_{\rm B} $ and an optical band gap of about 1.2\,eV. This 
is easily seen in the band structure shown in Fig.\ \ref{fig:res3}.  
Analysis of the corresponding partial densities of states reveals strong 
polarization of the vanadium atoms in the antiferroelectrically 
distorted chains, whereas polarization of the dimerizing vanadium atoms 
is almost negligible.


To conclude, band theory as based 
on density functional theory is well capable of correctly describing 
the metal-insulator transitions of $ {\rm VO_2} $ provided that the 
appropriate exchange-correlation functional is used. In particular, 
while LDA and GGA fail to account for the insulating behaviour of the 
low-temperature phases, the recently proposed hybrid functionals lead 
to a finite band gap for both the $ {\rm M_1} $ and $ {\rm M_2} $ 
phases and the antiferromagnetic ordering observed for the latter. 
In addition, much 
better agreement with photoemission data has been found. Our results 
put $ {\rm VO_2} $ in line with the metal-insulator system $ {\rm NbO_2} $ 
and metallic $ {\rm MoO_2} $, which both display the same structural 
distortions and have already been successfully described within density  
functional theory \cite{nbo2,moo2,moo2fs}. We are not aware of any other 
theory, which has provided a consistent and successful description of 
the structural distortions of the rutile-type early transition-metal 
dioxides as well as the MIT's of the $ d^1 $ members. Finally, the 
present study sheds new light on the old discussion about the 
predominant influence of either structural distortions or electronic 
correlations as no further interactions beyond those covered by density 
functional theory and the Hartree-Fock exchange have to be introduced. 
However, recent experimental studies indicate that the insulator-to-metal 
transition is initiated by rearrangements of the electronic system 
\cite{cao10,dachraoui11}. A theoretical understanding of these data is 
the subject of current research.

I am particularly grateful for vivid discussions and support to 
K.-H.\ H\"ock, T.\ Kopp, J.\ Mannhart, M.\ Marsman, P.\ Saxe, 
P.\ C.\ Schmidt, M.\ Stephan, and E.\ Wimmer. 
The calculations were performed using Material Design's MedeA 
computational environment as well as the computational facilities 
of Materials Design. 
This work was supported by the DFG through TRR 80 and by the BMBF 
through project 03SF0353B.

\end{document}